\documentclass[twocolumn,amsmath,amssymb]{revtex4}

\usepackage{graphicx}
\usepackage{bm}

\begin{document}

\title{A symmetrical method to obtain shear moduli from microrheology}

\author{Kengo Nishi\textsuperscript{1}}
\author{Maria L. Kilfoil\textsuperscript{2,3}}
\author{Christoph F. Schmidt\textsuperscript{1,4}}
\author{F. C.\ MacKintosh\textsuperscript{5,6,7,8,9}}
\address{
\textsuperscript{1}Third Institute of Physics-Biophysics, University of G\"{o}ttingen, 37077 G\"{o}ttingen, Germany\\
\textsuperscript{2}Alentic Microscience Inc., Halifax, NS B3H 0A8, Canada\\
\textsuperscript{3}Department of Physics, University of Massachusetts Amherst, Amherst, MA 01003, USA\\
\textsuperscript{4}Department of Physics, Duke University, Durham, NC 27708, USA\\
\textsuperscript{5}Department of Chemical \& Biomolecular Engineering, Rice University, Houston, TX 77005, USA\\
\textsuperscript{6}Center for Theoretical Biological Physics, Rice University, Houston, TX 77030, USA\\
\textsuperscript{7}Department of Chemistry, Rice University, Houston, TX 77005, USA\\
\textsuperscript{8}Department Physics \& Astronomy, Rice University, Houston, TX 77005, USA\\
\textsuperscript{9}Department of Physics and Astronomy, Vrije Universiteit, 1081HV Amsterdam, The Netherlands}

\date{\today}

\begin{abstract}
Passive microrheology typically deduces shear elastic loss and storage moduli from displacement time series or mean-squared displacement (MSD) of thermally fluctuating probe particles in equilibrium materials. Common data analysis methods use either Kramers-Kronig (KK) transformations or functional fitting to calculate frequency-dependent loss and storage moduli. We propose a new analysis method for passive microrheology that avoids the limitations of both of these approaches. In this method, we determine both real and imaginary components of the complex, frequency-dependent response function $\chi(\omega) = \chi^{\prime}(\omega)+i\chi^{\prime\prime}(\omega)$ as direct integral transforms of the MSD of thermal particle motion. This procedure significantly improves the high-frequency fidelity of $\chi(\omega)$ relative to the use of KK transformation, which has been shown to lead to artifacts in $\chi^{\prime}(\omega)$. We test our method on both model data and experimental data. Experiments were performed on  solutions of worm-like micelles and dilute collagen solutions. While the present method agrees well with established KK-based methods at low frequencies, we demonstrate significant improvement at high frequencies using our symmetric analysis method, up to almost the fundamental Nyquist limit.
\end{abstract}


\maketitle

\section{Introduction}
Bulk viscoelastic properties of soft materials are usually measured by mechanical rheometers, in which, for instance, the shear response of a material is measured under an applied oscillatory torsion ~\cite{Ferry,Bird,Larson,Doi,Graessley}. 
Complementary \emph{microrheological} techniques have been developed over the past two decades that permit the measurement of the local viscoelastic response of soft materials on microscopic length scales and at much higher frequencies than are possible with conventional rheometers~\cite{MasonWeitz_PRL_1995,microrheology:mason_wietz97,MRlaser:Gittes97,MRlaser:Schnurr97,MRreview}.  
In passive microrheology, the linear response characteristics of a material are inferred from the thermal motion of embedded probe particles. Provided that the medium is in thermal equilibrium, the fluctuation-dissipation theorem (FDT) \cite{FDToriginal,Kubo} implies that one can completely determine the linear response of a probe particle from the thermal fluctuations of that particle. Furthermore, provided that this micromechanical response can be accurately modeled in terms of macroscopic transport properties such as the viscosity $\eta$ or the viscoelastic shear modulus $G(\omega)$, e.g., using the Stokes formula or its generalizations, microrheology can be used to measure such bulk viscoelastic properties~\cite{MasonWeitz_PRL_1995,MRlaser:Gittes97,MRlaser:Schnurr97,MRreview,LevineLubensky:PRL}. 

The practical implementation of passive microrheology is subject to at least three limitations: (1) the temporal bandwidth and spatial resolution limits of the method used to measure the fluctuations, (2) the artifacts introduced by the analysis of these fluctuations to derive the micromechanical probe-particle response, and (3) the accuracy and appropriateness of models such as generalizations of the Stokes formula that are employed to relate the particle response to bulk material properties. 
Here, we address primarily the second of these limitations by introducing an improved method of analysis to obtain the local, micromechanical compliance or response function $\chi$ of the probe particle from displacement time-series data. Our analysis presupposes that either a direct displacement time series $x(t)$ or the mean-squared displacement (MSD) $\langle\left(x(t'+t)-x(t')\right)^2\rangle$ of particle motion is measured over about three decades or more in $t$. Various experimental techniques can be used for such measurements. The MSD of an ensemble of probe beads can, for instance, be obtained using light scattering methods, including dynamic light scattering (DLS) and diffusing wave spectroscopy (DWS), where the intensity fluctuations of the coherently scattered light can be related to the MSD~\cite{Microrheology:chapter}. These methods offer high spatial resolution and a wide bandwidth ($\sim$ 1 nm and $10-10^5$ Hz for DWS). Moreover, the averaging extends over hundreds or more probe particles, resulting in good statistics. This averaging, however, can become problematic if particles reside in inhomogeneous micromechanical environments.  

Alternatively, laser interferometry can be used to track thermal motions of single beads~\cite{MRlaser:Gittes97,MRlaser:Schnurr97,MasonWirtzKuo_PRL_1997,bfp}. This method delivers $\sim$ 1 nm spatial resolution and a frequency range of 0.1 Hz up to 100's of kHz, and averaging over multiple probe particles must be done sequentially. 
Finally, individual bead motions can be imaged directly over time using video microscopy, employing standard optical microscopes and digital cameras. The high-frequency limit is determined by the camera, and is typically less than 100 Hz, but can be extended up to 10s or even 100s of kHz with the use of specialized high-speed cameras ~\cite{liu:2006}. In this approach, 10s to 100s of beads can be tracked simultaneously, resulting in good statistics of the ensemble-averaged MSD. 
Importantly, both of the latter tracking methods retain single-bead trajectories for further evaluation. Among other things, this can be used to directly identify inhomogeneities in the micromechanical environments of the probe particles. 
From the displacement time series $x(t)$, one calculates the power spectral density (PSD), typically by fast Fourier transformation (step (a) in Fig.\ 1), or the mean-squared displacement (MSD) (step (b) in Fig.\ 1). 
\begin{figure}[h]
\centering
\includegraphics[height=10cm]{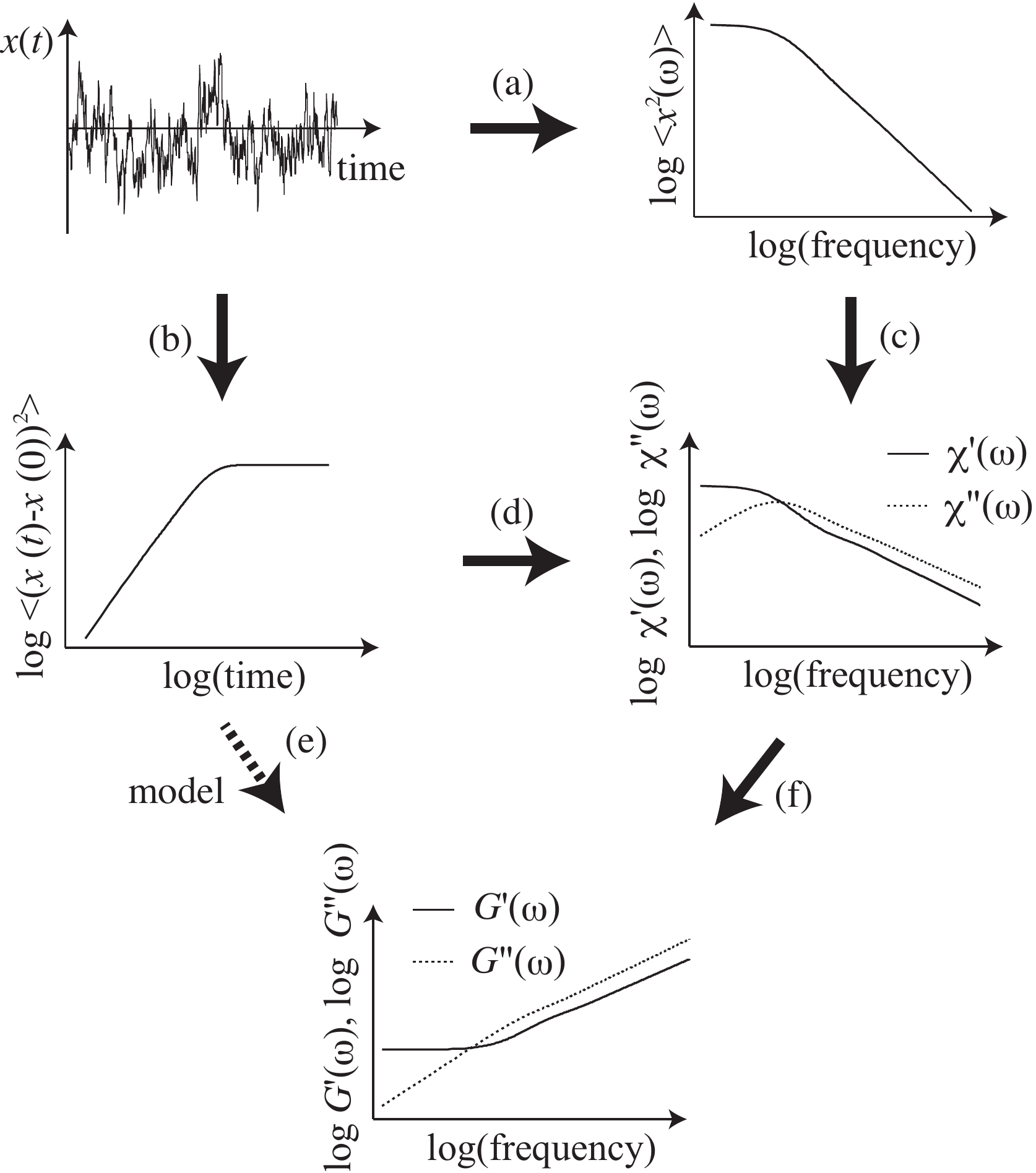}
\caption{Schematic of common data analysis procedures employed in passive microrheology. From the time series $x(t)$, one can calculate the displacement power spectral density (PSD) or the mean-squared displacement (MSD), as indicated by steps (a) and (b), respectively. 
From the PSD, one can determine the response function $\chi$ by direct application of the FDT, as indicated in step (c). This step, however, usually requires the use of a Kramers-Kronig integral transformation, the accuracy of which is limited by the bandwidth of the PSD. The shear modulus can then can be determined via the generalized Stokes formula (f). Alternatively, one can estimate the shear modulus from a functional fit to the Laplace transform of the MSD (e). In the latter, the result depends on the assumed functional form for fitting.}
  \label{fgr:example}
\end{figure}

The MSD and PSD are entirely equivalent, as they are simply related by Fourier transformation. Thus, either of these can be obtained (indicated by steps (a) and (b) in Fig.\ 1) to a level of accuracy and over a time/frequency range limited by the noise and detection bandwidth.  
From the PSD, one can directly determine the imaginary part $\chi''$ of the response function via the FDT (Fig.\ 1 step (c)). A significant advantage of this approach is its rigor up to this point: the accuracy of the result is limited only by experimental noise and detection band-width. A significant practical limitation, however, of prior implementations of this method enters with the calculation of the real part of the response function $\chi^{\prime}$, which is needed to determine the full complex shear modulus. This has led to a loss of accuracy at high frequencies over as much as a decade below the Nyquist limit, which imposes a cut-off in the required Kramers-Kronig integral transformation. 

A common alternative approach starts with the calculation of the MSD of particle motion (Fig.\ 1 step (b)). One can determine the complex shear modulus in an approximate manner by performing a Laplace transform on the MSD (Fig.\ 1 (e)) and fitting with an assumed functional form. One can then transform this fitted function to obtain an estimate of the frequency-dependent shear modulus using the generalized Stokes formula. Even assuming that the generalized Stokes formula is valid, the result can only be as good as the assumed functional form for the fitting. The choice of that function is either empirical or based on an expected form of the shear modulus. It thus represents an uncontrolled approximation: functions that look very similar in the Laplace domain can represent very different functions when continued to the Fourier (frequency) domain. In practice, the approach often depends on some knowledge or expectation of the rheological properties of the medium. 

Our aim here is to develop a more direct and still rigorous and objective approach to the calculation of the local micromechanical compliance $\chi(\omega)$ directly from the MSD (Fig.\ 1 (d)). From this, the complex shear modulus can also be determined directly (Fig.\ 1 (f)), provided the response can be accurately modeled in terms of the macroscopic rheology, e.g., via a Stokes-like formula. In section II, we review the relevant linear response theory and present our analysis method, applying it to simulated data, constructed for a  viscoelastic medium. In sections III and IV we present experimental results, comparing our new method with other approaches. Finally, in section V we discuss the limits of our approach and compare with alternative approaches.

\section{Background and Theory\label{sec:theory}} 
\subsection{Linear response\label{sec:linresp}} 
The simplest and most common approaches to micromechanical characterization measure the response of some time-dependent observable $A(t)$ to an applied force or field $f(t)$ that perturbs an otherwise equilibrium system. Provided this force is small enough, the response to be assumed to be linear in the force. In that case, the response can be expressed as \cite{Chandler,Chaikin:Lubensky}
\begin{equation}
\langle A(t)\rangle=\int_{-\infty}^\infty \chi(t-t')f(t')dt', \label{Chidef}
\end{equation}
where the response function $\chi(t)$ must vanish for $t<t'$, since the response cannot anticipate the force. Thus, the upper limit of the convolution integral above can be taken to be $t$. The average $\langle\cdots\rangle$ refers to an ensemble average with the same, time-dependent forcing $f(t)$. In the following, this average is implied, and we will drop the explicit notation. 

For a spherical particle of radius $a$ moving in a Newtonian liquid of viscosity $\eta$, the natural observable $A$ of interest is the particle velocity $v(t)$, which responds to a mechanical force $f(t)$ that is conjugate to the particle position $x$. On time scales long enough to neglect inertial effects, Eq.\ \eqref{Chidef} reduces to an instantaneous response $v=\mu f$, where $\mu=1/(6\pi\eta a)$ is the particle mobility or inverse drag coefficient. Here, we assume that the Stokes formula describes the drag force on the particle. This limit corresponds to $\chi(t)=\frac{1}{6\pi\eta a}\delta(t)$ above. Similarly, in the limit of a purely elastic response with no inertia, Eq.\ \eqref{Chidef} reduces to an instantaneous response in the particle position $x(t)\propto f(t)$. For a particle in an incompressible elastic medium with shear modulus $G$, the Stokes formula generalizes to $\chi(t)=\frac{1}{6\pi G a}\delta(t)$ above. 
More generally, by Fourier transforming this linear response relation in Eq.\ \eqref{Chidef}, we obtain 
$\tilde x(\omega)=\tilde\chi(\omega)\tilde f(\omega)$, where 
\begin{equation}
\tilde\chi=\frac{1}{6\pi G(\omega) a},\label{Stokes}
\end{equation} 
and $G(\omega)=G'(\omega)-iG''(\omega)$ is the complex, frequency-dependent shear modulus. 
Thus, if $\chi(t)$ or its Fourier transform $\tilde\chi(\omega)$ can be measured, then one can, in principle, determine $G(\omega)$. 

\subsection{Fluctuation-Dissipation Theorem \label{sec:theory}}
A key prediction of linear response theory is that the way a system returns to equilibrium following a small perturbation is governed by the same dynamics as the thermal fluctuations of the system about equilibrium. This is the essence of Onsager's regression hypothesis, which was proven from first principles in the Fluctuation-Dissipation Theorem \cite{FDToriginal,Kubo,Chandler}. 
Here, we consider the position $x$ of a probe particle that responds to a (perturbing) force $f$, which is thermodynamically conjugate to $x$. 
In the absence of the perturbation $f$, the position $x(t)$ fluctuates in time, in a way most naturally described by the correlation function
\begin{equation}
K(t)=\left< x(t)x(0)\right>,\label{Kdef}
\end{equation}
which we assume can be measured over a wide dynamic range of delay times $t$. Importantly, this function is strictly even. 
As long as linear response holds, e.g., as long as the perturbing force is sufficiently small, the coordinate $x(t)$ is linearly related to the perturbing force $f(t)$ via a response function $\chi(t)$:
\begin{equation}
x(t)=\int_{-\infty}^t \chi(t-t')f(t')dt'. \label{Chidef1}
\end{equation}

The Fluctuation-Dissipation Theorem (FDT) states that the correlation function $K$ and the response function $\chi$ are related by \cite{Chandler}
\begin{equation}
 kT\chi(t)=-\frac{d}{dt}K(t) \qquad{\mbox{for $t\ge0$}}.\label{FDT}
\end{equation}
For $t<0$, we take $\chi(t)=0$ in order to both enforce causality (no response in anticipation of the applied force) and 
render the integral in Eq.\ \eqref{Chidef} a full convolution. 

In view of the strictly even symmetry of $K(t)$ and odd symmetry of its derivative in Eq.\ \eqref{FDT}, it is convenient to define the even and odd functions
\begin{equation}
\chi_{E,O}(t)=\frac{1}{2}\left[ \chi(t) \pm \chi(-t) \right].
\end{equation}
The factor of 2 ensures that $\chi(t)=\chi_E(t)+\chi_O(t)$. 
It is always possible to decompose a function into a sum of even and odd functions. 
But, since $\chi(t)$ is only nonzero for $t>0$, it is also true that $\chi(t)=2\chi_E(t)=2\chi_O(t)$ for $t > 0$. 
Since the correlation function $K$ must be even, its derivative and the right-hand side in the FDT relation in Eq.\ \eqref{FDT} above must be odd. Thus, while the relation in Eq.\ \eqref{FDT} between $\chi$ and $K$ is only valid for $t>0$, it can be extended to all times using $\chi_O(t)$
\begin{equation}
2kT\chi_O(t)=-\frac{d}{dt}K(t)\qquad{\mbox{for all $t$}}.\label{ChiO}
\end{equation}
We can Fourier transform (FT) both sides of Eq.\ \eqref{ChiO}, resulting in
\begin{equation}
2kT\,\tilde{\chi}_O(\omega)=i\omega\tilde{K}(\omega)\,.\label{eq:FtFTD}
\end{equation}
Here, $\tilde{K}(\omega)$ is the Fourier transform of $K(t)$, which is also the power spectral density (PSD) of $x(t)$:
\begin{equation}
C(\omega)=\int_{-\infty}^\infty e^{i\omega t}K(t)\,dt\,.
\end{equation}

Given that $\chi(t)=\chi_{E}(t)+\chi_{O}(t)$ and that the FT of an even (odd) function is purely real (imaginary), it follows that
\begin{equation}
\tilde{\chi}=\chi'(\omega)+i\chi''(\omega),
\end{equation}
where $i\chi''(\omega)=\tilde{\chi}_O(\omega)$. Thus, the FDT can also be stated as
\begin{equation}
2kT\,\chi''(\omega)=\omega C(\omega)\,.\label{eq:FDT}
\end{equation}
From the power spectrum $C(\omega)$ and from $\chi''(\omega)$, one can determine $\chi'(\omega)=\tilde{\chi}_E(\omega)$, and therefore the full $\tilde{\chi}$ using a Kramers-Kronig transformation~\cite{MRlaser:Gittes97,MRlaser:Schnurr97,MRreview}
\begin{eqnarray}
\chi^{\prime}(\omega)&=&\frac{1}{\pi}\int_{-\infty}^{\infty}\frac{(\xi+\omega)\chi^{\prime\prime}(\xi)}{\xi^2-\omega^2}\,d\xi \label{KK}\\
&=&\frac{2}{\pi}\int_0^\infty\cos(\omega t)\,dt\int_0^\infty\chi^{\prime\prime}(\xi)\sin(\xi t)\,d\xi\,.\nonumber
\end{eqnarray}
While this is a valid method for obtaining $\tilde{\chi}$, it suffers from limitations due to the unavoidable truncation of the integrals in Eq.\ \eqref{KK}, due to the finite bandwidth with which $x(t)$ or $C(\omega)$ are measured in experiments. In practice, in typical microrheology applications, while $\chi^{\prime\prime}(\omega)$ can be determined with the same bandwidth as the PSD, $\chi^{\prime}(\omega)$
determined from Eq.\ \eqref{KK} is severely distorted over approximately one decade below the fundamental Nyquist limit. This fundamental asymmetry in obtaining real and imaginary parts of $\chi$ results in errors for both $G'(\omega)$ and $G''(\omega)$ using 
Eq.\ \eqref{Stokes}.

\subsection{Symmetric Approach}
In the following, we develop an alternative method for obtaining $\tilde\chi(\omega)$ and $G(\omega)$ from either the correlation function of probe particle displacement or (equivalently) from their mean-squared displacement (MSD). We first define the following explicitly odd function
\begin{equation}
K_O(t)=
    \left\{ \begin{array}{cc}
       K(t) & t > 0 \,\, \\
       -K(t) & t < 0\,,
    \end{array}\right. 
\end{equation}
Note that this function has a discontinuity at $t = 0$, while its derivative is continuous at that point. Following the reasoning leading to Eq.\ \eqref{ChiO}, one can see that
\begin{equation}
2kT\chi_E(t)=-\frac{d}{dt}K_O(t)\qquad{\mbox{for all $t$}}.\label{ChiE}
\end{equation}
It is tempting to Fourier transform both sides of Eq.\ \eqref{ChiE}, as in Eq.\ \eqref{eq:FtFTD}. The aforementioned discontinuity in $K_O(t)$ would, however, lead to a pole in $\tilde{K}_O(\omega)$ of strength proportional to $K(0)$, which may become ill-defined for viscoelastic media, for which the MSD can become unbounded. One can avoid the discontinuity by taking the derivative appearing on the right hand side of Eq.\ \eqref{ChiE} before Fourier transforming. As noted above, the derivative of $K_O$ is continuous at $t=0$. 

Alternatively, since $K(t)$ is related to the MSD, $M(t)$, by
\begin{equation}\label{eq:K_t_to_MSD}
M(t)=\left< (x(t)-x(0))^2 \right>=2[K(0)-K(t)]\,,
\end{equation}
one can define an odd function $M_O(t)$ as
\begin{equation}
M_O(t)=
    \left\{ \begin{array}{cc}
       M(t) & t > 0 \nonumber \\
       -M(-t) & t < 0,\nonumber
    \end{array}\right. 
\end{equation} 
which is continuous at $t=0$. 
With this definition, it follows from Eq.\ \eqref{ChiE} that 
\begin{equation}
4kT\chi_E(t)=\frac{d}{dt}M_O(t)\qquad{\mbox{for all $t$}}.\label{ChiEM}
\end{equation}
Now the real part of the response function $\chi'(\omega)$ can be obtained either from
\begin{equation}
2kT\chi'(\omega)=-i\omega \tilde M_O(\omega)/2=\omega\int^\infty_0 \sin(\omega t)M(t)\,dt\label{ChiM1}
\end{equation}
or 
\begin{equation}
2kT\chi'(\omega)=\int^\infty_0 \cos(\omega t)\dot M(t)\,dt,\label{ChiM}
\end{equation}
where the dot indicates the time derivative. 
The latter is more appropriate in practice, especially for an unbounded MSD. For instance, in a viscoelastic fluid with $M(t)\sim t^z$ for $0<z<1$, $\dot M\rightarrow0$ as $t\rightarrow\infty$ and $\chi^{\prime}\sim\omega^{-z}$. 
In any case, with a finite bandwidth, Eq.\ \eqref{ChiM} is likely preferable to Eq.\ \eqref{ChiM1}, since $\dot M(t)$ should vanish for large $t$ for all but Newtonian liquids, making the integral well-behaved at large $t$. The corresponding expression for $\chi^{\prime\prime}$ in terms of $\dot M(t)$ is
\begin{equation}
2kT\chi^{\prime\prime}(\omega)=\int_0^{\infty}\sin(\omega t)\,\dot M(t)\, dt.\label{ChiMp}
\end{equation}
The combination of Eqs.\ \eqref{ChiM} and \eqref{ChiMp} represent a symmetric derivation of $\chi'$ and $\chi^{\prime\prime}$, in contrast to the use of a Kramers-Kronig transformation. This should improve the high-frequency results for $G'(\omega)$ and $G''(\omega)$ for viscoelastic materials.

\subsection{Application to a viscoelastic material with known $M(t)$ and $\chi(t)$ \label{sec:analytical_results}}
To illustrate the performance of the new method using Eqs.~\eqref{ChiM} and \eqref{ChiMp}, we simulated microrheology data for a network of semiflexible filaments, where $G^{\prime}(\omega),\,G^{\prime\prime}(\omega)\sim \omega^{3/4}$ at high frequencies~\cite{MRlaser:Gittes97,MRlaser:Schnurr97,1998_Gittes_PRE}. 
Following Ref.~\cite{1998_Gittes_PRE}, we generated an exact MSD for a probe particle embedded in such a network, $M(t)$, up to a multiplicative constant, as a sum over thermal bending modes of the filaments making up the network: 
\begin{eqnarray}
M(t)&=&\sum_{n=1}^{\infty}\frac{1}{n^4}\left( 1-e^{-n^4|t|} \right)\,,\quad t \ge0\,.
\end{eqnarray}
Here, $n$ is the mode number and time is measured in units of the longest relaxation time for $n=1$. Because this function converges well for $n \geq 11$, we terminated the summation at $n = 11$. We sampled $M(t)$ at a frequency of 160 (in units of the inverse of the longest relaxation time). This expression for $M(t)$ can then be used to obtain $\chi^{\prime}({\omega})$ and $\chi^{\prime\prime}({\omega})$, following Eqs.~\eqref{ChiM} and \eqref{ChiMp}. 
The result for $\chi'$ is shown in Fig.\ \ref{fgr:example} (Symmetric Method), along with the exact $\chi'(\omega)$ obtained for this model, using 
\begin{eqnarray}\label{eq:chi_omega_WLC}
2kT\chi(\omega)&&=\sum_{n=1}^\infty \frac{1}{n^4-i\omega} \nonumber \\
&&=\sum_{n=1}^\infty \frac{n^4}{n^8+\omega^2}+i\sum_{n=1}^\infty \frac{\omega}{n^8+\omega^2} \nonumber \\
&&=\chi^{\prime}(\omega)+i\chi^{\prime\prime}(\omega)\quad .
\end{eqnarray}
As above, the series was truncated at $n=11$.

\begin{figure}[t]
\centering
  \includegraphics[height=6cm]{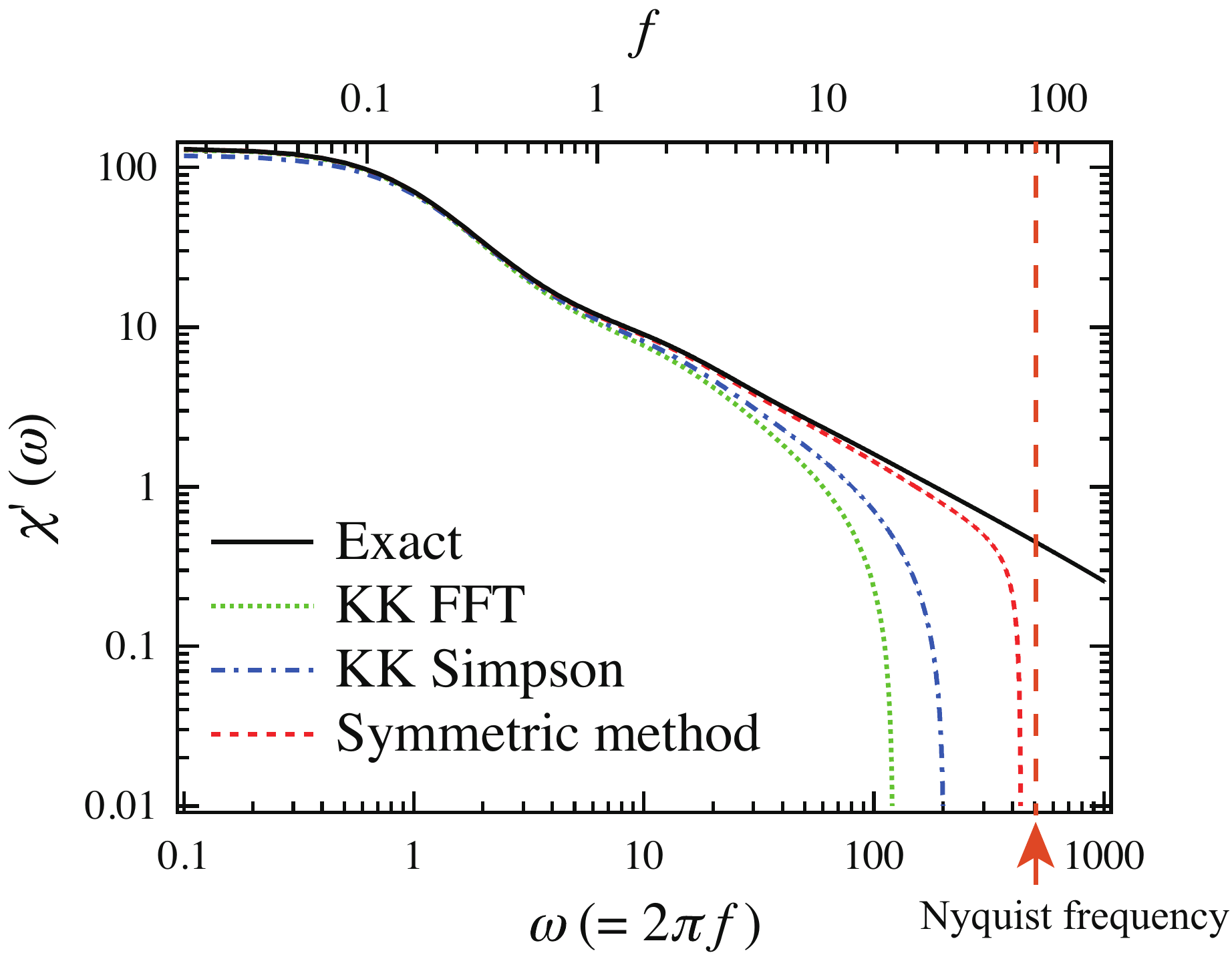}
  \caption{Calculation of $\chi^{\prime}(\omega)$ from simulated mean-squared displacement data: Exact $\chi^{\prime}(\omega)$ for idealized semiflexible polymer network, together with $\chi^{\prime}(\omega)$ obtained from sampled exact expressions for PSD (KK FFT and KK Simpson) or from the numerical derivative and exact integration of $M(t)$ (Symmetric Method) for the semiflexible gel. The corresponding Nyquist frequency is shown as the vertical red dashed line.}
  \label{fgr:example}
\end{figure}

Here, the derivative $\dot M$ was computed numerically using the five-point stencil method~\cite{Fornberg}, with which the derivative can be obtained to order $\delta^4$ in the spacing $\delta$ between consecutive data points. This is more accurate than a simple (order $\delta^2$) local slope of $M(t)$ obtained from pairs of consecutive data points. Specifically, we use
\begin{widetext} 
\begin{equation}
\dot M(t)=\frac{-M(t+2\delta)+8M(t+\delta)-8M(t-\delta)+M(t-2\delta)}{12\delta} + \mathcal(O)(\delta^4).
\end{equation} 
For the first two time points, we have used
\begin{equation}
\dot M(t)=\frac{-25M(t)+48M(t+\delta)-36M(t+2\delta)+16M(t+3\delta)-3M(t+4\delta)}{12\delta} + \mathcal(O)(\delta^4).
\end{equation} 
\end{widetext} 
To increase the accuracy of the subsequent integral, we also applied Simpson's rule, which is a Newton-Cotes formula for approximating the integrand $f(t)$ using quadratic polynominals, resulting in
\begin{equation}
\int_{x_0}^{x_2}dx\, f(x)\approx\frac{\delta}{3}\left(f(x_0)+4f(x_1)+f(x_2)\right)
\end{equation}
for equally spaced points $x_0$, $x_1$, and $x_2$, separated by a distance $\delta$. 

In order to compare with prior analysis methods, we used the exact PSD for an idealized semiflexible polymer network, 
\begin{eqnarray}
C(\omega)&=&\int_{-\infty}^{\infty}dt \quad K(t)e^{i\omega t}=\sum_{n=1}^\infty\frac{1}{n^8+\omega^2}\,,
\end{eqnarray}
again truncating this summation at $n = 11$. This exact PSD was then used together with Eq.\ \eqref{eq:FDT} to obtain $\chi^{\prime\prime}({\omega})$ numerically, sampled up to the Nyquist frequency (80 in our units). The real part of the response function $\chi^{\prime}({\omega})$ was then calculated using a Kramers Kronig integral. The KK integral was evaluated over a finite frequency range \cite{MRlaser:Schnurr97,landau_lifschitz_statphys2}. This can be done in either of two ways, as indicated in Fig.\ \ref{fgr:example}. In the first of these (KK FFT), the KK integral was calculated as the convolution of the functions $1/\omega$ and $\chi^{\prime\prime}(\omega)$ as shown in the first line of Eq.\ \eqref{KK}. To speed this calculation, we performed additional Fourier and inverse Fourier transformations using fast Fourier transformations (FFT) in the last line of Eq.\ \eqref{KK}. In the second method (KK Simpson), $\chi^{\prime}$ is obtained with an improved integration algorithm for Eq.\ \eqref{KK} using Simpson's rule.  This provides, at the cost of speed, a substantial  improvement over simple FFT, particularly at high frequencies; although both KK-based methods still exhibit high frequency artifacts due to finite bandwidth. In contrast, we observe improvement in the high-frequency region when avoiding the Kramers-Kronig integral altogether, using instead the new symmetric method described above.

\section{Experimental \label{sec:experiment}}

\subsection{Materials}
Wormlike micelles were prepared from the surfactant cetylpyridinium chloride (CPyCl) dissolved in $0.5\,\mathrm{M}$ NaCl in purified water, with strongly binding counterions, sodium salicylate (NaSal). CPyCl and NaSal were obtained from Sigma Aldrich Corp.\ (St. Louis, MO, USA). In this study, all samples had a molar ratio Sal/CPy = 0.5 and the concentration of CPyCl was kept at $2\,\mathrm{wt}\%$. The micellar solutions were stored at a controlled temperature of 30$^\circ$C, which was above the Krafft point of this system~\cite{Buchanan_PRE2005}.

Collagen type I (rat tail) was purchased as a stock solution with a concentration of 11.32 mg/ml in 20 mM acetic acid (BD Biosciences). For the experiments, type I collagen was diluted to the desired final concentrations of 0.57 mg/mL in water.
A small quantity ($< 0.001\%$ by volume) of COOH surface-modified polystyrene beads with a diameter of 1 or $2\,\mu\mathrm{m}$ (Kisker Biotech GmbH $\&$ Co. KG) was added to each sample before measurement.
 
\subsection{Methods}
All microrheology experiments were performed on a custom-built optical microscope equipped with optical traps essentially as described earlier~\cite{Buchanan_PRE2005}. Briefly, an infrared laser ($\lambda=1064$ nm, NdVO$_4$, COMPASS, Coherent Inc., Santa Clara, CA) was coupled into the sample via the microscope objective lens (Zeiss, Neofluar, 100$\times$, NA = 1.3) using immersion oil ($n_{\rm oil}$=1.5, Cargille LTD, Cedar Grove, NJ). This was used to trap particles with a typical power in the sample of 15 - 40 mW. Back-focal plane laser interferometry was used for precise position detection of the probe particles~\cite{1998_Gittes_OptLett}. The lateral ($x$ and $y$) displacements of the trapped particle were detected with a quadrant photodiode (YAG444-4A, Perkin Elmer, Vaudreuil, Canada). The current signals of the QPD were amplified by a low-noise analog differential amplifier (custom built) and sampled at the desired frequency, maximally 200 kHz, by a FPGA A/D board (NI PXI-7833R, National Instruments, Austin, TX, USA). The time series voltage data were converted to displacements in nm using independently measured calibration factors obtained from particles from the same batch trapped in water~\cite{Atakhorrami_PRE2006}.  

\section{Results}
\begin{figure}[h]
\centering
 \includegraphics[height=6cm]{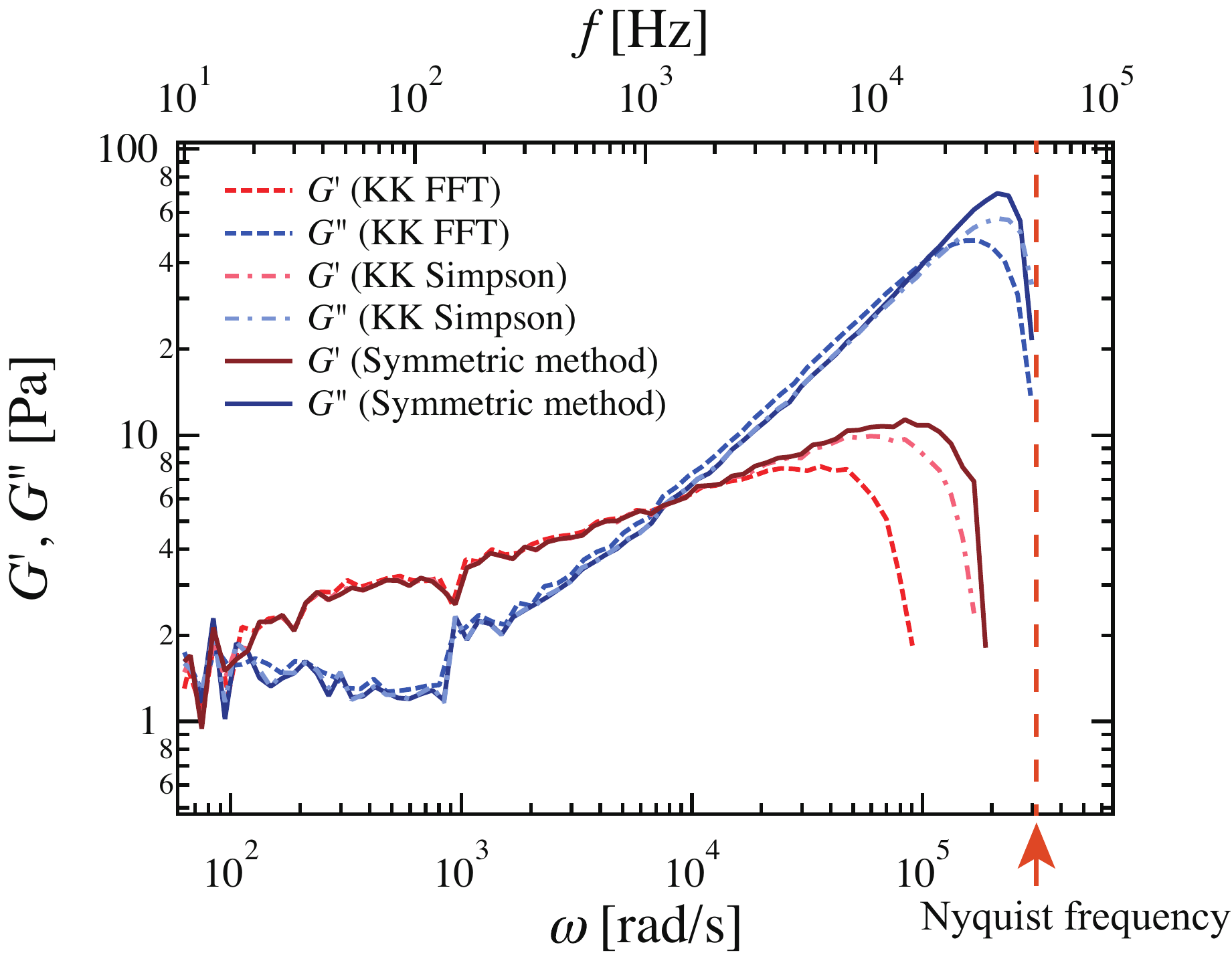}
  \caption{Complex shear moduli for a 2 wt$\%$ wormlike micelle solution, calculated from the primary time-series data using the three methods described in the the text. Data were smoothed by logarithmic binning for plotting. The corresponding Nyquist frequency is shown as the vertical dashed red line.}
  \label{fgr:example}
\end{figure}

To illustrate the difference between the new symmetric approach and the KK integral method, we recorded displacement fluctuation data for suspensions of 1 $\mu$m polystyrene beads in 2 wt$\%$ wormlike micelle solutions, using the custom-built instrument described above. The sampling rate was 100 kHz and the total recording time was 10 s. In order to minimize the elastic confinement by the optical trap, we used a low laser power of $\sim40$ mW, which resulted in a trap stiffness of $\sim$ 100 $\mu$N/m measured in buffer. The trap produced an apparent added elastic modulus $G^{\prime}\sim1$ Pa in the final result for the shear moduli. The trap effect was corrected for by subtracting this constant~\cite{Atakhorrami_PRE2006}.

As shown in Fig.~3, the KK FFT method becomes unreliable at high frequencies, roughly a decade below the Nyquist frequency. The KK Simpson method gives already a significantly more accurate result up to about half the Nyquist frequency. The present symmetric method proves to be even more robust in the high frequency region, where the KK methods fail. Experimental noise in the ACF or the MSD is a potential issue when performing the discrete derivative of either time series to evaluate $\chi$($\tau$). Random experimental noise, however, is strongly repressed for short times or high frequencies because there are good statistics for the first several points of the ACF or the MSD, and the resulting functions appear smooth. The improvement obtained by the new symmetric method is significant because high bandwidth in the measurement of viscoelastic response is one of the main merits of microrheology techniques.

\begin{figure}[h]
\centering
\includegraphics[height=6cm]{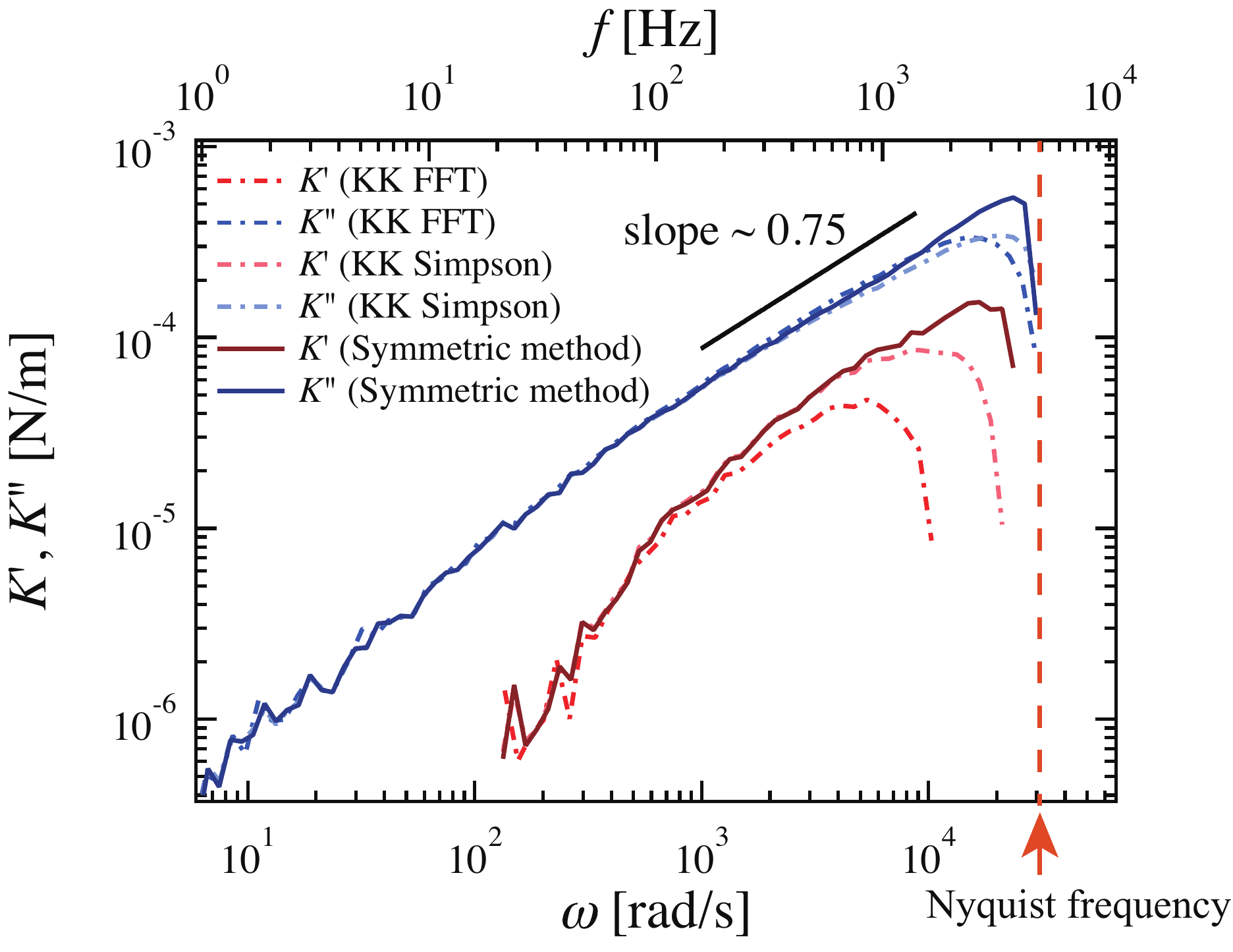}
  \caption{Complex spring constant  $K(\omega)=K^{\prime}(\omega)-iK^{\prime\prime}(\omega)=1/\chi(\omega)$ measured for a 0.57 mg/ml collagen solution evaluated with the three different calculation methods. Data were smoothed by logarithmic binning for plotting. The corresponding Nyquist frequency is shown as the vertical dashed red line.}
  \label{fgr:example}
\end{figure}

Microrheology data were also recorded for collagen solutions with a concentration of 0.57 mg/ml. The bead diameter was 2 $\mu\mathrm{m}$, the sampling rate was 10 kHz, and the measuring time was 100 s. The filaments in the solution were fibrillar collagen aggregates with a diameter of $\sim200$ nm and contour lengths of tens of $\mu\mathrm{m}$, as described in the literature~\cite{MR_collagen_Forde_PLOSOne_2013}, rather than single collagen triple helices (diameter $\sim$ 2 $\mu \mathrm{m}$). In such a network with inhomogeneity on large scales, the Stokes formula for the particle response is unlikely to be applicable: 
on the scale of the probe particles we use, we cannot assume that the material can be treated accurately as a viscoelastic continuum. Nevertheless, once the solution reaches thermal equilibrium, we are justified in assuming linear response theory to relate the position fluctuations of a single probe particle to the local micromechanical compliance $\chi(\omega)$. 
Hence, instead of the complex shear modulus $G(\omega)$, we report our results in terms of a complex spring constant $K(\omega)$, defined as $K(\omega)=K^{\prime}(\omega)-iK^{\prime\prime}(\omega)=1/\chi(\omega)$. 

Fig.\ 4 shows the results for a dilute collagen solution. Although the micromechanical compliance is not necessarily directly related to the macroscopic shear modulus, its frequency dependence is similar to that of the complex shear modulus measured in semiflexible filament networks that can be approximated as viscoelastic continua. The new symmetric method again is clearly less affected by the finite bandwidth of recording than the KK Simpson method in the high-frequency regime, and performs much better than the KK FFT method, across the entire frequency range. In particular, the high-frequency regime of $K^{\prime}(\omega)$, calculated by the symmetric method, shows power law scaling for more than a decade in frequency with a slope of $\sim0.75$.  This scaling corresponds to theoretical predictions for both semiflexible polymer fluctuations and the rheological response of semiflexible networks~\cite{Amblard,Morse,1998_Gittes_PRE}. 

\section{Conclusions}
We have developed a new analysis method for passive microrheology data, using the FDT. In this approach, we determine the real and imaginary components of the complex, frequency-dependent response function $\chi(\omega)$ as direct integral transforms of the mean-squared displacement (MSD) of thermal particle motion. The approach treats real and imaginary part of the response function in a more symmetric way than do methods using KK transformations~\cite{MRlaser:Gittes97,MRlaser:Schnurr97,MRreview}, which reduces the effect of finite-bandwidth recording on $\chi^{\prime}(\omega)$. 
We illustrate the performance of the symmetric method by comparing various methods applied to numerically generated model data. 
We also performed microrheology experiments in solutions of WLM and collagen, which demonstrate both the practical implementation of our method, as well as the substantial improvement in resolving the viscoelastic response at high frequency. 

The method presented here is most readily implemented for viscoelastic or solid-like media, for which the MSD is grows sub-linearly or is bounded in time. For media exhibiting long-time liquid-like or viscous behavior, the integrals in Eqs.\ (\ref{ChiM},\ref{ChiMp}) can become ill-defined. In practice, this is not a problem for optical tweezers-based microrheology, since trapping can be used to regularize these integrals at long times. Moreover, the main point of the method introduced here is to address problems in resolving short-time or high frequency behavior. Thus, other methods can be used to resolve the low-frequency response. Interestingly, for liquid-like response, Yanagishima et al.\ have demonstrated a method for determining the rheology by analysis of the velocity autocorrelation function, which also avoids the use of KK transformations \cite{Frenkel2011}. This was shown to be particularly effective for viscous fluids.

\appendix
\section{Background}
An exact expression for $K(t)$ for the case of a semiflexible gel may be obtained from \cite{1998_Gittes_PRE}
\begin{eqnarray}
K(t)&\propto&\sum_n\int e^{-i\omega\,t}\frac{1}{n^4+\omega^2}\nonumber\\
&=&\operatorname{Re} \sum_n\frac{1}{n^2}   \int\cos(\omega\,t)\frac{1}{n^2-i\omega}\nonumber\\
&=&\frac{1}{2}\sum_{n=1}^\infty\frac{1}{n^4}e^{-|t|n^4}\,,\quad t \ge0\,.
\end{eqnarray}

An exact expression for $C(\omega)$ (power spectral density) for the case of a semiflexible gel may be obtained from Eq.~\eqref{eq:chi_omega_WLC}, which may be re-written as 
\begin{equation}
\chi({\omega})\propto\sum_{n=1}^\infty\frac{n^4+i\omega}{n^8+\omega^2}\,,
\end{equation}
with imaginary part
\begin{equation}
\chi^{\prime\prime}({\omega})\propto\sum_{n=1}^\infty\frac{\omega}{n^8+\omega^2}\,.
\end{equation}
From Eq.~\ref{eq:FDT}, we obtain 
\begin{equation}
C(\omega)\propto\sum_{n=1}^\infty\frac{1}{n^8+\omega^2}\,.
\end{equation}

\begin{acknowledgments}
M.L.K. was supported by NSF (OISE-1444209). F.C.M.\ was supported in part by NSF (PHY-1427654). The research leading to these results has received funding from the European Research Council under the European Union's Seventh Framework Programme (FP7/2007- 2013)/ERC grant agreement no.\ 340528 (C.F.S and K.N.), and the DFG Collaborative Research Center SFB 937 (Project A2) (C.F.S.). M.L.K., C.F.S.\ and F.C.M.\ thank the Kavli Institute for Theoretical Physics, where a portion of this work was done. 
\end{acknowledgments}


\begin{thebibliography}{29}
\expandafter\ifx\csname natexlab\endcsname\relax\def\natexlab#1{#1}\fi
\expandafter\ifx\csname bibnamefont\endcsname\relax
  \def\bibnamefont#1{#1}\fi
\expandafter\ifx\csname bibfnamefont\endcsname\relax
  \def\bibfnamefont#1{#1}\fi
\expandafter\ifx\csname citenamefont\endcsname\relax
  \def\citenamefont#1{#1}\fi
\expandafter\ifx\csname url\endcsname\relax
  \def\url#1{\texttt{#1}}\fi
\expandafter\ifx\csname urlprefix\endcsname\relax\def\urlprefix{URL }\fi
\providecommand{\bibinfo}[2]{#2}
\providecommand{\eprint}[2][]{\url{#2}}

\bibitem[{\citenamefont{Ferry}(1980)}]{Ferry}
\bibinfo{author}{\bibfnamefont{J.~D.} \bibnamefont{Ferry}},
  \emph{\bibinfo{title}{Viscoelastic Properties of Polymers}}
  (\bibinfo{publisher}{Wiley}, \bibinfo{year}{1980}).

\bibitem[{\citenamefont{Bird}(1987)}]{Bird}
\bibinfo{author}{\bibfnamefont{R.~B.} \bibnamefont{Bird}},
  \emph{\bibinfo{title}{Dynamics of Polymeric Liquids}}
  (\bibinfo{publisher}{Wiley}, \bibinfo{year}{1987}).

\bibitem[{\citenamefont{Larson}(1988)}]{Larson}
\bibinfo{author}{\bibfnamefont{R.~G.} \bibnamefont{Larson}},
  \emph{\bibinfo{title}{Constitutive Equations for Polymer Melts and
  Solutions}} (\bibinfo{publisher}{Butterworths}, \bibinfo{year}{1988}).

\bibitem[{\citenamefont{Doi and Edwards}(1988)}]{Doi}
\bibinfo{author}{\bibfnamefont{M.}~\bibnamefont{Doi}} \bibnamefont{and}
  \bibinfo{author}{\bibfnamefont{S.~F.} \bibnamefont{Edwards}},
  \emph{\bibinfo{title}{The Theory of Polymer Dynamics}}
  (\bibinfo{publisher}{Clarendon Press}, \bibinfo{year}{1988}).

\bibitem[{\citenamefont{Graessley}(1974)}]{Graessley}
\bibinfo{author}{\bibfnamefont{W.~W.} \bibnamefont{Graessley}},
  \emph{\bibinfo{title}{The Entanglement Concept in Polymer Rheology}}
  (\bibinfo{publisher}{Springer-Verlag}, \bibinfo{year}{1974}).

\bibitem[{\citenamefont{Mason and Weitz}(1995)}]{MasonWeitz_PRL_1995}
\bibinfo{author}{\bibfnamefont{T.~G.} \bibnamefont{Mason}} \bibnamefont{and}
  \bibinfo{author}{\bibfnamefont{D.~A.} \bibnamefont{Weitz}},
  \bibinfo{journal}{Phys. Rev. Lett.} \textbf{\bibinfo{volume}{74}},
  \bibinfo{pages}{1250} (\bibinfo{year}{1995}).

\bibitem[{\citenamefont{Mason et~al.}(1997{\natexlab{a}})\citenamefont{Mason,
  Gang, and Weitz}}]{microrheology:mason_wietz97}
\bibinfo{author}{\bibfnamefont{T.}~\bibnamefont{Mason}},
  \bibinfo{author}{\bibfnamefont{H.}~\bibnamefont{Gang}}, \bibnamefont{and}
  \bibinfo{author}{\bibfnamefont{D.}~\bibnamefont{Weitz}}, \bibinfo{journal}{J.
  Opt. Soc. Am.} \textbf{\bibinfo{volume}{14}}, \bibinfo{pages}{139}
  (\bibinfo{year}{1997}{\natexlab{a}}).

\bibitem[{\citenamefont{Gittes et~al.}(1997)\citenamefont{Gittes, Schnurr,
  Olmsted, MacKintosh, and Schmidt}}]{MRlaser:Gittes97}
\bibinfo{author}{\bibfnamefont{F.}~\bibnamefont{Gittes}},
  \bibinfo{author}{\bibfnamefont{B.}~\bibnamefont{Schnurr}},
  \bibinfo{author}{\bibfnamefont{P.~D.} \bibnamefont{Olmsted}},
  \bibinfo{author}{\bibfnamefont{F.~C.} \bibnamefont{MacKintosh}},
  \bibnamefont{and} \bibinfo{author}{\bibfnamefont{C.~F.}
  \bibnamefont{Schmidt}}, \bibinfo{journal}{Phys. Rev. Lett.}
  \textbf{\bibinfo{volume}{79}}, \bibinfo{pages}{3286} (\bibinfo{year}{1997}).

\bibitem[{\citenamefont{Schnurr et~al.}(1997)\citenamefont{Schnurr, Gittes,
  MacKintosh, and Schmidt}}]{MRlaser:Schnurr97}
\bibinfo{author}{\bibfnamefont{B.}~\bibnamefont{Schnurr}},
  \bibinfo{author}{\bibfnamefont{F.}~\bibnamefont{Gittes}},
  \bibinfo{author}{\bibfnamefont{F.~C.} \bibnamefont{MacKintosh}},
  \bibnamefont{and} \bibinfo{author}{\bibfnamefont{C.~F.}
  \bibnamefont{Schmidt}}, \bibinfo{journal}{Macromolecules}
  \textbf{\bibinfo{volume}{30}}, \bibinfo{pages}{7781} (\bibinfo{year}{1997}).

\bibitem[{\citenamefont{MacKintosh and Schmidt}(1999)}]{MRreview}
\bibinfo{author}{\bibfnamefont{F.~C.} \bibnamefont{MacKintosh}}
  \bibnamefont{and} \bibinfo{author}{\bibfnamefont{C.~F.}
  \bibnamefont{Schmidt}}, \bibinfo{journal}{Curr. Opin. Colloid Interface Sci.}
  \textbf{\bibinfo{volume}{4}}, \bibinfo{pages}{300} (\bibinfo{year}{1999}).

\bibitem[{\citenamefont{Callen and Welton}(1951)}]{FDToriginal}
\bibinfo{author}{\bibfnamefont{H.}~\bibnamefont{Callen}} \bibnamefont{and}
  \bibinfo{author}{\bibfnamefont{T.}~\bibnamefont{Welton}},
  \bibinfo{journal}{Physical Review} \textbf{\bibinfo{volume}{83}},
  \bibinfo{pages}{34} (\bibinfo{year}{1951}).

\bibitem[{\citenamefont{Kubo}(1966)}]{Kubo}
\bibinfo{author}{\bibfnamefont{R.}~\bibnamefont{Kubo}},
  \bibinfo{journal}{Reports on the Progress of Physics}
  \textbf{\bibinfo{volume}{29}}, \bibinfo{pages}{255} (\bibinfo{year}{1966}).

\bibitem[{\citenamefont{Levine and Lubensky}(2000)}]{LevineLubensky:PRL}
\bibinfo{author}{\bibfnamefont{A.}~\bibnamefont{Levine}} \bibnamefont{and}
  \bibinfo{author}{\bibfnamefont{T.}~\bibnamefont{Lubensky}},
  \bibinfo{journal}{Phys. Rev. Lett.} \textbf{\bibinfo{volume}{85}},
  \bibinfo{pages}{1774} (\bibinfo{year}{2000}).

\bibitem[{\citenamefont{Gardel et~al.}(2005)\citenamefont{Gardel, Valentine,
  and Weitz}}]{Microrheology:chapter}
\bibinfo{author}{\bibfnamefont{M.}~\bibnamefont{Gardel}},
  \bibinfo{author}{\bibfnamefont{M.}~\bibnamefont{Valentine}},
  \bibnamefont{and} \bibinfo{author}{\bibfnamefont{D.}~\bibnamefont{Weitz}},
  \emph{\bibinfo{title}{Microscale Diagnostic Techniques}}
  (\bibinfo{publisher}{Springer Verlag}, \bibinfo{year}{2005}), chap.
  \bibinfo{chapter}{Microrheology}.

\bibitem[{\citenamefont{Mason et~al.}(1997{\natexlab{b}})\citenamefont{Mason,
  Ganesan, van Zanten, Wirtz, and Kuo}}]{MasonWirtzKuo_PRL_1997}
\bibinfo{author}{\bibfnamefont{T.~G.} \bibnamefont{Mason}},
  \bibinfo{author}{\bibfnamefont{K.}~\bibnamefont{Ganesan}},
  \bibinfo{author}{\bibfnamefont{J.~H.} \bibnamefont{van Zanten}},
  \bibinfo{author}{\bibfnamefont{D.}~\bibnamefont{Wirtz}}, \bibnamefont{and}
  \bibinfo{author}{\bibfnamefont{S.~C.} \bibnamefont{Kuo}},
  \bibinfo{journal}{Phys. Rev. Lett.} \textbf{\bibinfo{volume}{79}},
  \bibinfo{pages}{3282} (\bibinfo{year}{1997}{\natexlab{b}}).

\bibitem[{\citenamefont{Gittes and Schmidt}(1998{\natexlab{a}})}]{bfp}
\bibinfo{author}{\bibfnamefont{F.}~\bibnamefont{Gittes}} \bibnamefont{and}
  \bibinfo{author}{\bibfnamefont{C.~F.} \bibnamefont{Schmidt}},
  \bibinfo{journal}{Opt. Lett.} \textbf{\bibinfo{volume}{23}},
  \bibinfo{pages}{7} (\bibinfo{year}{1998}{\natexlab{a}}).

\bibitem[{\citenamefont{Liu et~al.}(2006)\citenamefont{Liu, Gardel, Kroy, Frey,
  Hoffman, Crocker, Bausch, and Weitz}}]{liu:2006}
\bibinfo{author}{\bibfnamefont{J.}~\bibnamefont{Liu}},
  \bibinfo{author}{\bibfnamefont{M.}~\bibnamefont{Gardel}},
  \bibinfo{author}{\bibfnamefont{K.}~\bibnamefont{Kroy}},
  \bibinfo{author}{\bibfnamefont{E.}~\bibnamefont{Frey}},
  \bibinfo{author}{\bibfnamefont{B.}~\bibnamefont{Hoffman}},
  \bibinfo{author}{\bibfnamefont{J.}~\bibnamefont{Crocker}},
  \bibinfo{author}{\bibfnamefont{A.}~\bibnamefont{Bausch}}, \bibnamefont{and}
  \bibinfo{author}{\bibfnamefont{D.}~\bibnamefont{Weitz}},
  \bibinfo{journal}{Phys. Rev. Lett.} \textbf{\bibinfo{volume}{96}},
  \bibinfo{pages}{118104} (\bibinfo{year}{2006}).

\bibitem[{\citenamefont{Chandler}(1987)}]{Chandler}
\bibinfo{author}{\bibfnamefont{D.}~\bibnamefont{Chandler}},
  \emph{\bibinfo{title}{Introduction to Modern Statistical Mechanics}}
  (\bibinfo{publisher}{{O}xford {U}niversity {P}ress}, \bibinfo{year}{1987}).

\bibitem[{\citenamefont{Chaikin and Lubensky}(2000)}]{Chaikin:Lubensky}
\bibinfo{author}{\bibfnamefont{P.}~\bibnamefont{Chaikin}} \bibnamefont{and}
  \bibinfo{author}{\bibfnamefont{T.}~\bibnamefont{Lubensky}},
  \emph{\bibinfo{title}{Principles of Condensed Matter Physics}}
  (\bibinfo{publisher}{Cambridge University Press}, \bibinfo{year}{2000}).

\bibitem[{\citenamefont{Gittes and MacKintosh}(1998)}]{1998_Gittes_PRE}
\bibinfo{author}{\bibfnamefont{F.}~\bibnamefont{Gittes}} \bibnamefont{and}
  \bibinfo{author}{\bibfnamefont{F.~C.} \bibnamefont{MacKintosh}},
  \bibinfo{journal}{Phys. Rev. E} \textbf{\bibinfo{volume}{58}},
  \bibinfo{pages}{R1241} (\bibinfo{year}{1998}).

\bibitem[{\citenamefont{Fornberg}(1988)}]{Fornberg}
\bibinfo{author}{\bibfnamefont{B.}~\bibnamefont{Fornberg}},
  \bibinfo{journal}{Math. Comp.} \textbf{\bibinfo{volume}{51}},
  \bibinfo{pages}{699} (\bibinfo{year}{1988}).

\bibitem[{\citenamefont{Landau and
  Lifshitz}(1980)}]{landau_lifschitz_statphys2}
\bibinfo{author}{\bibfnamefont{L.~D.} \bibnamefont{Landau}} \bibnamefont{and}
  \bibinfo{author}{\bibfnamefont{E.~M.} \bibnamefont{Lifshitz}},
  \emph{\bibinfo{title}{Statistical Physics Pt. 2}}
  (\bibinfo{publisher}{Pergamon Press}, \bibinfo{address}{Oxford},
  \bibinfo{year}{1980}), \bibinfo{edition}{2nd} ed.

\bibitem[{\citenamefont{Buchanan et~al.}(2005)\citenamefont{Buchanan,
  Atakhorrami, Palierne, MacKintosh, and Schmidt}}]{Buchanan_PRE2005}
\bibinfo{author}{\bibfnamefont{M.}~\bibnamefont{Buchanan}},
  \bibinfo{author}{\bibfnamefont{M.}~\bibnamefont{Atakhorrami}},
  \bibinfo{author}{\bibfnamefont{J.~F.} \bibnamefont{Palierne}},
  \bibinfo{author}{\bibfnamefont{F.~C.} \bibnamefont{MacKintosh}},
  \bibnamefont{and} \bibinfo{author}{\bibfnamefont{C.~F.}
  \bibnamefont{Schmidt}}, \bibinfo{journal}{{P}hys. {R}ev. {E}}
  \textbf{\bibinfo{volume}{72}}, \bibinfo{pages}{011504}
  (\bibinfo{year}{2005}).

\bibitem[{\citenamefont{Gittes and
  Schmidt}(1998{\natexlab{b}})}]{1998_Gittes_OptLett}
\bibinfo{author}{\bibfnamefont{F.}~\bibnamefont{Gittes}} \bibnamefont{and}
  \bibinfo{author}{\bibfnamefont{C.~F.} \bibnamefont{Schmidt}},
  \bibinfo{journal}{Opt. Lett.} \textbf{\bibinfo{volume}{23}},
  \bibinfo{pages}{7} (\bibinfo{year}{1998}{\natexlab{b}}).

\bibitem[{\citenamefont{Atakhorrami et~al.}(2006)\citenamefont{Atakhorrami,
  Sulkowska, Addas, Koenderink, Tang, Levine, , MacKintosh, and
  Schmidt}}]{Atakhorrami_PRE2006}
\bibinfo{author}{\bibfnamefont{M.}~\bibnamefont{Atakhorrami}},
  \bibinfo{author}{\bibfnamefont{J.~I.} \bibnamefont{Sulkowska}},
  \bibinfo{author}{\bibfnamefont{K.~M.} \bibnamefont{Addas}},
  \bibinfo{author}{\bibfnamefont{G.~H.} \bibnamefont{Koenderink}},
  \bibinfo{author}{\bibfnamefont{J.~X.} \bibnamefont{Tang}},
  \bibinfo{author}{\bibfnamefont{A.~J.} \bibnamefont{Levine}}, ,
  \bibinfo{author}{\bibfnamefont{F.~C.} \bibnamefont{MacKintosh}},
  \bibnamefont{and} \bibinfo{author}{\bibfnamefont{C.~F.}
  \bibnamefont{Schmidt}}, \bibinfo{journal}{{P}hys. {R}ev. {E}}
  \textbf{\bibinfo{volume}{73}}, \bibinfo{pages}{061501}
  (\bibinfo{year}{2006}).

\bibitem[{\citenamefont{Shayegan and
  Forde}(2013)}]{MR_collagen_Forde_PLOSOne_2013}
\bibinfo{author}{\bibfnamefont{M.}~\bibnamefont{Shayegan}} \bibnamefont{and}
  \bibinfo{author}{\bibfnamefont{N.~R.} \bibnamefont{Forde}},
  \bibinfo{journal}{PLOS ONE} \textbf{\bibinfo{volume}{8}},
  \bibinfo{pages}{e70590} (\bibinfo{year}{2013}).

\bibitem[{\citenamefont{Amblard et~al.}(1996)\citenamefont{Amblard, Maggs,
  Yurke, Pargellis, and Leibler}}]{Amblard}
\bibinfo{author}{\bibfnamefont{F.}~\bibnamefont{Amblard}},
  \bibinfo{author}{\bibfnamefont{A.~C.} \bibnamefont{Maggs}},
  \bibinfo{author}{\bibfnamefont{B.}~\bibnamefont{Yurke}},
  \bibinfo{author}{\bibfnamefont{A.~N.} \bibnamefont{Pargellis}},
  \bibnamefont{and} \bibinfo{author}{\bibfnamefont{S.}~\bibnamefont{Leibler}},
  \bibinfo{journal}{Physical Review Letters} \textbf{\bibinfo{volume}{77}},
  \bibinfo{pages}{4470} (\bibinfo{year}{1996}).

\bibitem[{\citenamefont{Morse}(1998)}]{Morse}
\bibinfo{author}{\bibfnamefont{D.~C.} \bibnamefont{Morse}},
  \bibinfo{journal}{Macromolecules} \textbf{\bibinfo{volume}{31}},
  \bibinfo{pages}{7044} (\bibinfo{year}{1998}).

\bibitem[{\citenamefont{Yanagishima et~al.}(2011)\citenamefont{Yanagishima,
  Frenkel, Kotar, and Eiser}}]{Frenkel2011}
\bibinfo{author}{\bibfnamefont{T.}~\bibnamefont{Yanagishima}},
  \bibinfo{author}{\bibfnamefont{D.}~\bibnamefont{Frenkel}},
  \bibinfo{author}{\bibfnamefont{J.}~\bibnamefont{Kotar}}, \bibnamefont{and}
  \bibinfo{author}{\bibfnamefont{E.}~\bibnamefont{Eiser}},
  \bibinfo{journal}{Journal of Physics: Condensed Matter}
  \textbf{\bibinfo{volume}{23}}, \bibinfo{pages}{194118}
  (\bibinfo{year}{2011}).

\end{thebibliography}

\end{document}